\documentclass{aa}  

\usepackage{graphicx}

\usepackage{txfonts}

\usepackage{soul}
\usepackage[normalem]{ulem}

\usepackage[usenames]{color}

\usepackage{natbib}

\usepackage{setspace}

\begin{document} 

   \title{The effect of JWST/NIRSpec data reduction on the retrieval of WASP-39b atmospheric properties}

   \titlerunning{Effect of data reduction process in transmission spectroscopy}
   \authorrunning{Roy-Perez et al.}

   \author{J. Roy-Perez
          \inst{1},
          S. Pérez-Hoyos
          \inst{1},
          N. Barrado-Izagirre
          \inst{1}
          \and
          H. Chen-Chen
          \inst{1}
          }

   \institute{Escuela de Ingeniería de Bilbao, Universidad del País Vasco (UPV/EHU), Bilbao, Spain\\
              \email{juan.roy@ehu.eus}
             }

   \date{Received 20 November 2025 / Accepted 3 February 2026}

  \abstract
   {The \textit{James Webb} Space Telescope (JWST) provides exoplanetary transit observations with an unprecedented spectral range coverage, offering exceptional data on exoplanet atmospheres. Nevertheless, the presence of systematics in the data reduction process introduces small, but significant uncertainties that propagate into atmospheric retrievals. Understanding how these reduction choices affect interpreted atmospheric properties is essential.}
   {We aim to quantify the impact of different JWST/NIRSpec PRISM data-reduction processes, as well as the relevance of saturation on the retrieved atmospheric properties of WASP-39b. We also assess whether or not these biases are comparable to those introduced by assumptions made in atmospheric modelling, particularly in the treatment of aerosol extinction. Compared with previous similar efforts, we use the highest number of alternative data reductions and discuss the role of the saturated spectral region, without adding any additional data from other sources or random noise, in contrast to previous works.}
   {We performed a series of nested-sampling Bayesian retrievals using MultiNest and forward models generated with the Planetary Spectrum Generator. We analysed six independently reduced NIRSpec/PRISM spectra and compare retrievals using the full wavelength range as well as versions excluding the saturated 0.69–1.91 $\mu$m region. We further tested the effect of including three different cloud-opacity parametrisations.}
   {Significant discrepancies arise among retrievals based on different calibrations, affecting key atmospheric parameters such as temperature, molecular abundances, and cloud opacity, often at a level exceeding one order of magnitude. Excluding the saturated region reduces the inter-pipeline dispersion; conversely, however, it also increases degeneracies between parameters. Differences introduced by data reduction are comparable in magnitude to those produced by distinct cloud-opacity models. Bayesian evidence systematically favours non-flat aerosol extinction, although the preferred spectral behaviour depends on the specific calibration used.}
   {Variations among JWST/NIRSpec data-reduction pipelines produce measurable and often substantial differences in retrieved atmospheric properties of WASP-39b. These biases are similar in scale to the uncertainties on modelling assumptions, underscoring the importance of robust and homogeneous calibration procedures. The results also confirm that JWST data possess the sensitivity required to probe aerosol spectral behaviour, although such constraints remain strongly dependent on the adopted data-reduction strategy.}

  \maketitle


\section{Introduction}\label{sec:Introduction}

The characterisation of exoplanets is one of the primary objectives of the \textit{James Webb} Space Telescope (JWST). Since its launch, JWST has observed more than a hundred exoplanets, with many additional observations in planning. Most of these are transit observations, where the data are analysed using transit spectroscopy to infer atmospheric properties \citep{Barstow2015}.

The observation of WASP-39b during the Early Release Science (ERS) programme demonstrated the potential of the space telescope. The planet, with a mass of $0.28 \, M_J$ and a radius of $1.27 \, R_J$, has been extensively studied by the scientific community since its discovery in 2011 by \cite{Faedi2011}. Its first transit spectrum was presented by \cite{Sing2016} using the Space Telescope Imaging Spectrograph (STIS) instrument onboard the \textit{Hubble} Space Telescope (HST). Based on the absorption features of $\rm{Na}$ and $\rm{K}$ observed in the data (also seen in ground-based observations from \citealp{Nikolov2016}), a number of studies \citep{Fisher2016, Heng2016, Barstow2017} investigated whether the planet has a clear atmosphere or it is covered by clouds, but no clear conclusion were obtained. The observations of the planetary transit in the near-infrared (NIR) with the Wide Field Camera 3 (WFC3) allowed \cite{Wakeford2018} to better constrain the abundance of $\rm{H_2O}$ in the atmosphere. These data undoubtedly showed that the atmosphere was cloudy and later analyses (e.g. \citealp{Tsiaras2018, Pinhas2019, Fisher2018, Pinhas2018} and \citealp{Kirk2019}) were instrumental in attempts to determine its characteristics, while also improving existing constraints on the abundance of water.

In July 2022, during the observations of the ERS programme, the scientific team measured the transmission spectra of WASP-39b using the NIRCam \citep{Ahrer2023ERS}, NIRISS \citep{Feinstein2023ERS}, NIRSpec PRISM \citep{Rustamkulov2023}, and NIRSpec G395H \citep{Alderson2023} instrumental modes. In addition to the previously detected $\rm{Na}$, $\rm{K}$, and $\rm{H_2O}$ abundances, these observations led to the detection of the expected species $\rm{CO_2}$ and $\rm{CO}$, as well as, unexpectedly, $\rm{SO_2}$ (see also \citealt{Ahrer2023ERSCO2, Grant2023, Tsai2023SO2}). The observations also showed an absorption feature that could be produced by $\rm{H_2S}$, but this still lacks a robust confirmation.

The data collected during the observations have been extensively used by different teams. Some of them focused their studies on the formation and evolution of the planet \citep{Louca2023, Khorshid2024}. Differences in the terminators have been clearly detected by \cite{Espinoza2024}, which has also motivated the study of the atmospheric transport of chemical species and clouds with global circulation models \citep{Tsai2023GCM}. The wide spectral range of the data has also motivated the study of aerosol properties \citep{Roy2025} that also include microphysical models \citep{Arfaux2024}.

However, it is also important to be aware of the differences between observations of the same object and, particularly, between their data processing. For example, \cite{Lueber2024} made a comparison of the retrieved atmospheric properties of the planet when using the data from different JWST instrument modes. Similarly, \cite{Davey2025} studied the effect of binning the spectroscopic observations on the atmospheric retrievals. The effect of the reduction process was studied for the first time in \cite{Constantinou2023} comparing only two data reductions from NIRSpec instrument, and by \cite{Powell2024} for MIRI observations considering three different data reductions. Recently, \cite{Schleich2025} also studied the effect of random noise in the processed spectra of NIRSpec observations. These works show the strong influence that the observation and reduction process may have on the retrieved characteristics of a given planet. However, the number of alternative calibration procedures has been increasing steadily without a comprehensive recap of their expected impact in the retrievals.

Moreover, during the NIRSpec/PRISM observations, a region of the detector was partially saturated, as shown in \cite{Rustamkulov2023}. Each data reduction process published followed specific prescriptions to recover the lost information and safely use the whole spectral range for the atmospheric analysis. However, it is still unclear whether the different efforts to recover that information were fully successful or not and to which extent they affect the conclusions draw by previous works. Some works avoided the problem of the saturated region by using only a subset of the data (e.g. from 2.5 to 5 $\mu$m as in \citealt{Schleich2025}), occasionally adding data from other sources, such as HST to improve their results (see \citealt{Constantinou2023}). Again, there is no systematic analysis that compares each different approach and their results.

Thus, the purpose of the present work is to determine how different reduction processes applied to the same observations affect the retrieved atmospheric properties, even when they appear to be almost equivalent to each other. Here, we include a higher a number of reduced spectra than previous works, so the variability of atmospheric retrievals can be properly addressed. To complement \cite{Constantinou2023} and \cite{Schleich2025}, we chose to include the region below 3 microns as dealt with by each calibration process. This approach avoids introducing uncertainties derived from the calibration of other instruments and tests the self-consistency of the JWST data. We also aim to study the bias that the saturated region of the spectra introduces in the retrieved properties of the planet. To do so, we present a comparison of the results obtained when removing that specific region of the spectral data, without adding any additional sources of information.

Lastly, we try to put these reduction biases into context by comparing them with those introduced at the modelling stage; for instance, when using different cloud extinction models. Different works have claimed that JWST data has enough potential to constrain information about aerosols present on exoplanet atmospheres. Specifically, \cite{Constantinou2023, Lueber2024, Roy2025}, amongst others, have used different aerosol opacity definitions to study their role on WASP-39b transmission spectra, arguing that this increased degree of complexity is supported by the data. Thus, our aim is to provide more information on the properties of clouds present on the exoplanet, to complement the study of their effect on the retrieved atmosphere and check that the conclusion that non-flat extinction aerosols are supported by the JWST data still holds.

To do so, we have structured the text as follows. In Sect. \ref{sec:Data}, we present the set of spectral data used in this work. We briefly explore the different processing steps that can turn into different spectral characteristics for all of them. In Sect. \ref{sec:Methodology}, we show how we model the atmosphere of the planet and how we perform our retrievals based on Bayesian inference. In Sect. \ref{sec:Results}, we compare and discuss the retrieval differences obtained. In Sect. \ref{sec:Conclusions}, we summarise the main conclusions of this work and how they are related to previous works.

\section{Data} \label{sec:Data}

\begin{figure*}
\centering
\includegraphics[width=0.78\hsize]{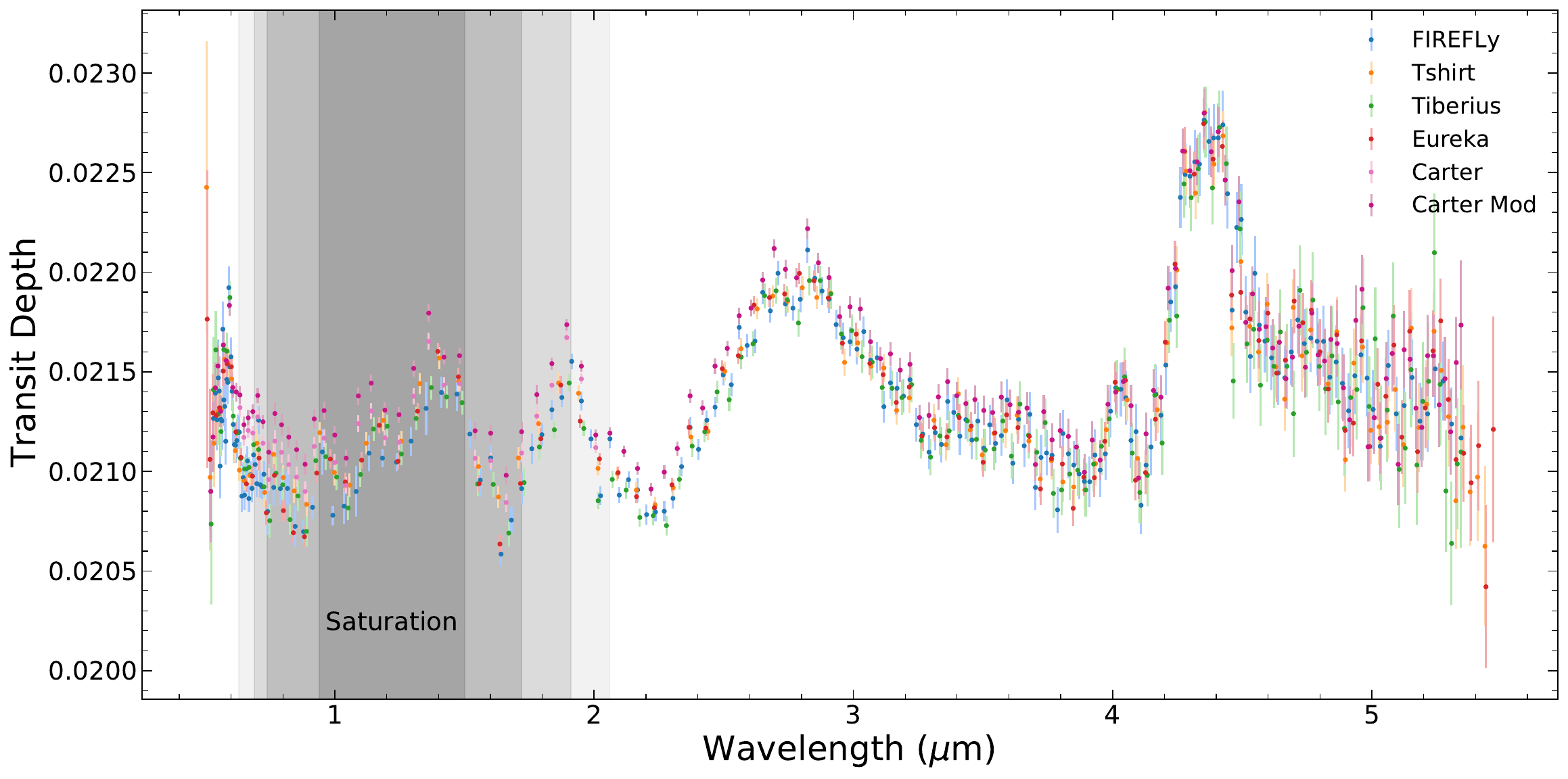}
\caption{Representation of the different spectral data used for the purpose of this work. The gray shades in the 0.63 - 2.06 micron range are related to the level of saturation: for the lightest region, only one group per integration was affected by saturation, while for the darkest region, up to four of the five groups were affected.}
          \label{FigEspecs}
\end{figure*}

The original data for the transit of WASP-39b used in this work correspond to the ERS programme of the JWST presented in \cite{Rustamkulov2023}. The 8.23 hour observations were taken on 10 July 2022 using the PRISM mode of the NIRSpec instrument. The ERS team used NIRSpec’s Bright Object Time Series (BOTS) mode with the NRSRAPID readout pattern, the S1600A1 slit (1.6" × 1.6") and the SUB512 sub-array. Throughout the exposure, 21,500 integrations were recorded, each with five 0.28 s groups up the ramp. As mentioned in the introduction, a region of the detector saturated during the observations. The position of the saturated pixels in the detector corresponds to a wavelength range between 0.63 and 2.06 $\mu$m. The saturated region is represented in Fig \ref{FigEspecs} as a gray shadowed area. However, the saturation was not homogeneous at all pixels and this is represented by the different gray shades used in the figure. For the lightest region, only one group per integration was affected by saturation, while for the darkest region, up to four of the five groups were affected. As mentioned in \cite{Rustamkulov2023}, the saturation produced a large point-to-point scatter of several thousand ppm in the transmission spectra. Applying various custom steps outside the regular data reduction process helped for potentially recovering the spectral information. \cite{Carter2024} performed a deeper analysis of this saturated region and discussed this correction in more detail.

\cite{Rustamkulov2023} presented four alternative calibrations of the original data, which we selected as a starting point for this work. They were obtained using four main reduction pipelines: FIREFLy \citep{RustamkulovFIREFLy}, Tshirt \citep{Tshirt}, Tiberius \citep{Kirk2017Tiberius, Kirk2021Tiberius}, and Eureka!+ExoTEP \citep{Eureka, BennekeExoTEP}. Henceforth, we refer to these transmission spectra by the name of the pipeline used for their extraction.

Later in 2024, \cite{Carter2024} reanalysed the observations of this transit. They did an analysis of the orbital and stellar parameters combining data taken by the four different instrumental modes of the JWST. From their work, we also take the reduced NIRSpec/PRISM spectrum, which we refer to in the following as `Carter'. An additional effort was made to recover the saturated region of the observations more effectively and accurately, which resulted in a spectral dependent modification of the affected part of the spectrum. We also include this modified version of the spectrum, hereafter referred to as `Carter-modified'.

At this point, it is important to note that the present work is not a comparison of the pipelines themselves. Its purpose is to evaluate how different data reduction procedures influence the atmospheric analysis results. Thus, we use the name of the pipelines just as a simpler way to name the different spectra (and not to evaluate the pipeline itself). It must be noted that beyond the chosen pipeline, the data reduction process consists of a series of steps, each one requiring its own input parameters. An interested reader is referred to original works from the ERS programme and the pipelines documentation. Here, we briefly summarise these steps to highlight possible differences that could lead to disparities in the final spectral data.

The process is divided in different stages. Detector corrections at group level are performed in Stage 1 for each integration. The bias subtraction, removal of bad pixels produced by cosmic rays or saturation, and/or the frequency-dependent noise correction are performed in this first stage. After cleaning the images, the flux for each integration is calculated fitting the ramp produced by its compounding groups. The wavelength calibration required to relate wavelength values to each pixel is executed in Stage 2. A second extraction of the background and outlier detection can be performed in Stage 3, this time at the integration level. At this third stage, the tracing of the source is identified to extract its spectrum from each integration, which are then used to generate the light curves of the transit. From this point on, each pipeline divides the remaining steps in a different number of stages, but they all perform similar processes. First, the spectral grid is defined to set the number of channels and, consequently, the binning for the final spectra. Then, the white light curve is computed. The next step consists of retrieving the orbital and stellar parameters fitting the white light curve with models of transits, stellar limb-darkening, and systematics. Once these parameters are obtained, the light curves of the different spectral channels are fitted separately. Lastly, the final spectrum is obtained from the planet sizes computed at each channel.

All these stages provide numerous opportunities to diverge during the reduction process. Some of them can be spotted even at the paper by \cite{Rustamkulov2023}. At the pixel level, there are many choices that can lead to dissimilarities in the extracted stellar spectra; for example, the chosen criteria for identifying a saturated pixel or a cosmic ray event or the pixel group selection for determining both the source trace and the background level. The selection of the channels for the spectral binning is also very relevant (see also \citealp{Davey2025}). Then, the choices made for fitting the models to the light curves extracted from the images can also affect heavily the retrieved planetary sizes. Different transit models are used for the retrievals and different systematics are considered among the different data reductions. As mentioned in \cite{Rustamkulov2023}, the limb darkening can have a huge impact on the extracted spectrum. This is the reason for all the reduction processes to fit it using the same parametrisation, but not retrieving always the same parameter values. In addition, when fitting the light curves of each wavelength channel, some reductions fix the limb darkening parameters to those obtained with the white light curve, while others fit them again.

Figure \ref{FigEspecs} shows a comparison of the six different spectra used in this work. At first glance, they all look very similar with particular differences mainly below the error bars. All of them capture the same shape of the planetary absorption throughout the spectral range. Not surprisingly, the largest difference can be seen in the saturated region, where Carter and Carter-modified spectra are able to retrieve more atmospheric absorption following their deeper analysis of the detector saturation. It is also relevant to highlight the different spectral grids chosen for each spectrum and this is not always with the same number of spectral points. Furthermore, there seems to be a noticeable dispersion between reductions at longer wavelengths, due to the lower brightness of the star at those wavelengths. It is also possible to appreciate a slight difference in the transit depth base reference value between data reductions. This simple vertical offset could be produced by the difference on the retrieved orbital parameters during the light curve fitting as highlighted in \cite{Rustamkulov2023}.

\begin{figure}
\centering
\includegraphics[width=0.9\hsize]{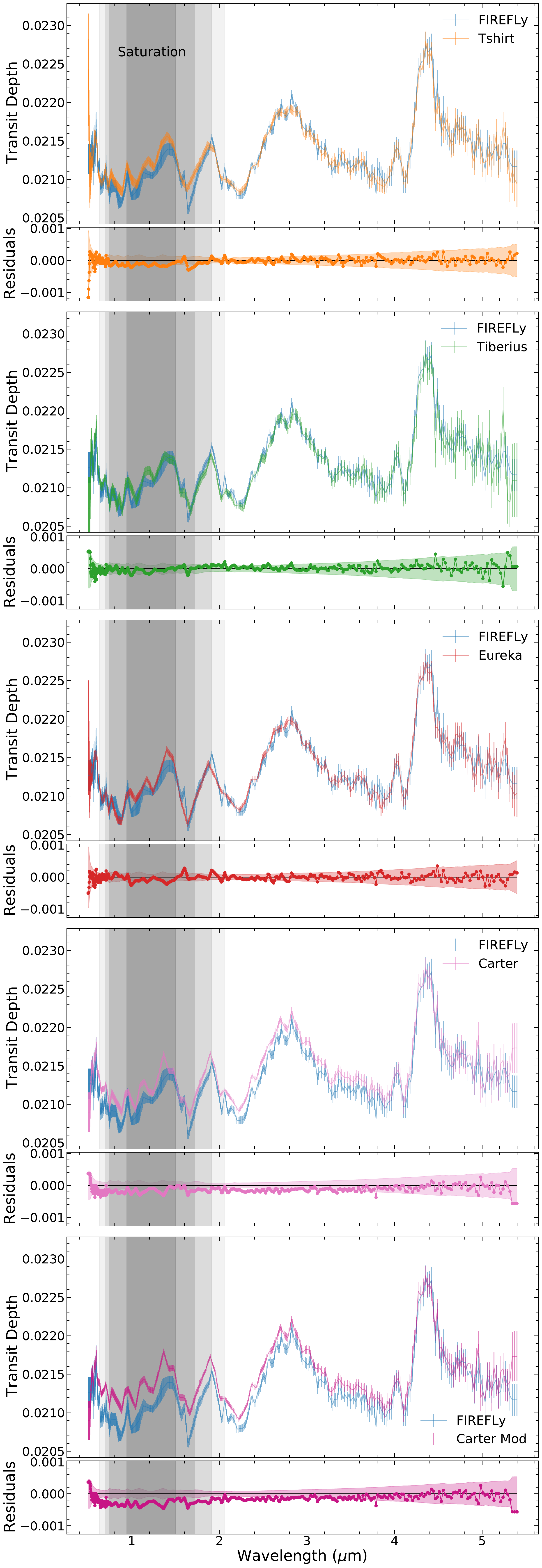}
\caption{Comparison of the spectra with the FIREFLy data reduction as a reference and their respective residuals. Coloured regions at the residuals represent error bar propagation. Grey regions show the saturated spectral range (0.63 - 2.06 $\mu$m) as in Fig. \ref{FigEspecs}.}
          \label{FigResiduals}
\end{figure}

Figure \ref{FigResiduals} shows a quantitative comparison of the data reductions used. We interpolated the spectra to a common spectral grid with resolving power of 200 and used the FIREFLy spectrum as reference. This choice was made simply for a direct comparison with the analysis from \cite{Rustamkulov2023}. For Tshirt, Tiberius, and Eureka, the residuals indicate that only a few peaks fall outside the error bars. Most of these peaks fall in the saturated regions, indicative of the different treatments of the saturated region of the detector. The spectra from \cite{Carter2024}, in addition to those peak differences, also show a clear offset that is not compatible with the error bars along most of the spectrum, possibly due to the additional offset introduced to cross-calibrate these data with other instruments.

\section{Methodology}\label{sec:Methodology}

The core idea of this work is to perform atmospheric retrievals using each of the calibrated spectra introduced in the preceding section. Comparing the retrieved parameters and their uncertainties provide information on how the data reduction process is affecting those retrievals and, hence, the specific parameters that will give robust results.

The first input we used were the spectra detailed in Sect. \ref{sec:Data}. However, as the saturated region could compromise the results and was sometimes removed, as a second step, we simply eliminated the saturated region to check how this change would affect the results. We removed the spectral points between 0.69 and 1.91 $\mu$m, forming the `persistently saturated region' defined in \cite{Carter2024}. It must be noted that, henceforth, there is no difference between the spectra labelled as 'Carter' and 'Carter-modified' as they are exactly the same outside the saturated region.

So far, only differences in the data have been proposed. We carried out an extra loop of simulations related to changes in the model atmosphere. With this last set of retrievals, we were able to compare the magnitude of the differences introduced in the retrieved atmospheric properties using both the data reduction and model aspects. While we used simple flat cloud absorption models for the first and second iterations, in the third, we used more complex cloud extinction models. This additional loop is intended to offer further insight and validation of the results presented in \cite{Roy2025}, helping to assess how sensitive cloud models are to data calibration.

\subsection{Tools}\label{subsec:Tools}

    The retrievals in the present work are based on the methodology described in \cite{Roy2025}. The forward simulations are generated with the Planetary Spectrum Generator (PSG; \citealt{PSG2018}). To generate the simulated spectra, we used the Planetary and Universal Model of Atmospheric Scattering (PUMAS; \citealt{PSGPUMAS, PSGPUMASScattering}), as the core radiative-transfer model, which uses a one-dimensional and plane-parallel description of the atmosphere. The spectra are computed with a resolving power of 200 using a correlated-k method. Higher resolutions were tested, but they produced negligible differences at the cost of significantly increasing the computation time. We used the line lists for $\rm{H_2O}$ \citep{Polansky2018}, $\rm{CO_2}$ \citep{Yurchenko2020}, $\rm{CO}$ \citep{Li2015}, $\rm{SO_2}$ \citep{Gordon2022}, $\rm{Na}$ \citep{Allard2019}, $\rm{H_2S}$ \citep{Gordon2022}, $\rm{CH_4}$ \citep{Yurchenko2024}, and $\rm{K}$ \citep{Allard2016}. Rayleigh scattering \citep{Maarten2005}, UV broad absorption, and collision-induced absorption by $\rm{H_2-H_2}$ \citep{Abel2011} and $\rm{H_2-He}$ \citep{Abel2012} are also included in the simulations. Multiple scattering is also taken into account by the radiative-transfer model.

    The retrievals were performed with MultiNest \citep{FerozMultiNest, BuchnerPyMultiNest}, a Bayesian inference tool based on nested sampling Monte Carlo algorithms \citep{Metropolis1953, Hastings1970, Skilling2006}. We used the Bayesian evidence $\ln Z$ for model comparison. When comparing two specific models, following Jeffrey's scale \citep{Jeffrey, BayesTrotta}, values of the Bayes factor, $\ln B_{1,2} = \ln Z_{1} - \ln Z_{2}$, that are greater than $5.0$ indicate that the first model is decisively better than the second, but differences greater than $1.0$ are enough to claim that one model is preferred. Instead, values lower than $1.0$ indicate that there is not enough evidence to claim which one is better. In this last case, Occam's razor reasoning is followed: the model with the lowest number of free parameters is favoured.

\subsection{Atmospheric description}\label{subsec:AtmDesc}
   
    Several parameters must be included in the retrievals to simulate the transit spectra. The size of the star, $R_{*}$, which determines the total amount of light reaching the planet and the planetary diameter, $D_{pl}$, which constrains the planetary surface blocking the star light, are included as free parameters with the tight prior Gaussian distribution constraints calculated in \cite{Faedi2011}: $179,000 \pm 7,000$ km (i.e. $1.27 \pm 0.04$ $R_J$) for $D_{pl}$ and $0.895 \pm 0.023 \, R_{\odot}$ for $R_{*}$. 

    To describe the atmosphere, we defined a log-uniform pressure profile with 40 levels between $10^{-6}$ and $10^1$ bar. We tested that increasing the grid to 100 vertical levels did not produce any significant differences, while increasing the computation time. Its vertical extension is described by the scale height, which depends on the planetary gravity, the atmospheric mean molecular weight and temperature \citep{Kreidberg2018Exoplanets}. For the planetary gravity, we assumed a Gaussian prior distribution of log $g = 0.63 \pm 0.05$ (with $g$ in $\mathrm{m} \, \mathrm{s}^{-2}$) in base to the calculations of \cite{Faedi2011}. Then, $D_{pl}$ and log $g$ are referenced to the pressure of 10 bar, which corresponds to the lowest layer of our model.

    Despite the fact that more complex temperature profiles were recently retrieved for WASP-39b (e.g. \citealt{Ma2025}, combining data from different JWST instruments), here we follow the discussion and calculations of \cite{Roy2025} and we used an isothermal profile. The rationale for doing so is that a model selection analysis comparing different vertical profiles based on the \cite{Kitzmann2020} parameterisation for NIRSpec/PRISM observations showed that the isothermal profile provided the highest Bayesian evidence; therefore, it was favoured against more sophisticated vertical profiles for this dataset. Similar results were obtained by \cite{Lueber2024}, which are also compatible with the results of \cite{Constantinou2023}. Therefore, we include the atmospheric temperature, $T_{iso}$, with a uniform prior distribution between 500 and 2000 K in our retrievals.

    Our model atmosphere is mainly composed of $\rm{H_2}$ and $\rm{He}$ at fixed abundances. Following the discussion from \cite{Roy2025}, we also included $\rm{H_2O}$, $\rm{CO_2}$ $\rm{CO}$, $\rm{SO_2}$, $\rm{Na}$, and $\rm{H_2S}$. The abundances are described as uniform profiles throughout the entire atmosphere and their volume mixing ratio is treated as a free parameter with uniform prior distributions between $10^{-12}$ and $10^{-1}$. We performed free chemistry retrievals where the chemical composition was allowed to vary without composition constraints. Thus, the value of the mean molecular weight is computed based on the retrieved chemical abundances.

\subsection{Cloud parametrisation}\label{subsec:Clouds}  

   Clouds largely determine the height at which the atmosphere becomes opaque in transit geometry. Following the results of the cloud profile model selection of \cite{Roy2025}, we used a uniform vertical profile to describe the abundance $X_{Cl}$ of aerosols. Similarly to what has been discussed on the thermal vertical profile, the work by \cite{Roy2025} discussed model selection using Bayesian evidence for a number of aerosol vertical profiles, which strongly favour a uniform vertical distribution. While this is most likely an oversimplification of the real situation, it is justified by the fact that the retrieval is only being sensitive to a reduced region of the atmosphere (for this particular dataset) and, hence, a more realistic vertical description is not supported by the information content.

   For the spectral behaviour of the aerosols, we used three different parametrisations of the extinction cross-section, $Q_{ext}$, which characterises how effectively the aerosol removes radiation from the incoming radiation beam. First, we used a simple flat parametrisation, where $Q_{ext}$ is constant with wavelength. When performing retrievals with this description, we fixed the value of $Q_{ext}$. Thus, the cloud opacity was determined only by its abundance, $X_{Cl}$. The second model of cloud extinction is an exponential dependence of the optical thickness with wavelength \citep{Angstrom} as in \cite{Roy2025}. In this case, the cloud opacity is defined by the free parameters $X_{Cl}$ and the $\textup{\r{A}}$ngström exponent ($\alpha$). This model has already been used in exoplanet atmospheric studies (see \citealt{Lecavelier2008, McDonald2017, Pinhas2019}; or \citealt{Barstow2020}). It has the advantage of a very simple parameterisation that mimics the behaviour of aerosols across a range of sizes: the smaller the value of $\alpha$, the bigger the aerosol it resembles. As shown in \cite{Roy2025}, this approach has the potential to be sensitive to the overall dependence of opacity with wavelength.

    Lastly, for a more realistic approach to aerosol opacity, we included a Mie parametrisation of the optical properties of the aerosol using the external database Modelled Optical Properties of enSeMbles of Aerosol Particles (MOPSMAP; \citealt{MOPSMAP}). We followed the assumptions made in \cite{Roy2025} and we defined the aerosols in our simulations as spherical particles with a log-normal size distribution. To describe this distribution, we used the effective radius, $r_{eff}$, and the effective variance, $v_{eff}$, which are the mean and deviation of a size distribution weighted by the cross-section of the particles \citep{Hansen}. For simplicity, we fixed $v_{eff}$ to $0.1$ during the retrieval calculations. The real and the imaginary part of the refractive index were fixed to $n_r = 1.4$ and $n_i = 0.0001$. \cite{Constantinou2023} assumed a number of possible compositions for the aerosols during their retrievals, based on thermochemical expectations \citep{Morley2013} and using the refractive index data listed in \cite{WakefordSing2015}. Similarly, assuming the composition of the aerosols as the estimated condensable species $\rm{MgSiO_3}$ and $\rm{SiO_2}$, \cite{Ma2025} found that using a combination of different homogeneous particle-size distributions is needed for an optimal fit to the combined data of the different JWST instruments. However, this adds an extra layer of complexity that could mask other features in the $Q_{ext}$ investigation that we propose. Furthermore, there is a long list of potential thermochemical candidates even for the solar system clouds that have not been detected so far, particularly in the giant planets, possibly masked by the complex photochemistry above the cloud decks \citep{Irwin2009Book}. For this reason, we stuck to a simpler approach disregarding the cloud composition, namely, the dependence of refractive indices with wavelength, but taking into account the role of particle size in the opacity.

   Thus, in this last parameterisation, the $Q_{ext}$ dependency on wavelength is only due to the particle-size distributions and not to the absorption properties of the aerosols. Therefore, when using this parametrisation, the cloud opacity calculated during retrievals is defined by the free parameters $X_{Cl}$ and $r_{eff}$. Table \ref{tprioris} summarises the prior probability distributions introduced in MultiNest for every free parameter included in the simulations.

    \begin{table}[]
    \caption{Prior probability distributions for the free parameters included in the retrievals.}
    \label{tprioris}
    \centering
    \begin{spacing}{1.2}
    \begin{tabular}{lccc}
    \hline\hline
    Parameter & Distribution & Range & Units       \\ \hline
    
    $D_{pl}$    & Gaussian    & $179,000 \pm 7,000$   & km      \\
    $R_{*}$     & Gaussian    & $0.895 \pm 0.023$   &   $R_{\odot}$ \\
    $log \, g$  & Gaussian    & $0.63 \pm 0.05$     & $\mathrm{m} \, \mathrm{s}^{-2}$ \\
    $T_{iso}$  & Uniform   & [$500.0$, $2000.0$]   & K      \\ 
    $X_i$  & Log-uniform   & [$10^{-15}$, $10^{-1}$]   & -     \\ \hline
    \textit{Cloud extinction} & & & \\
    $\alpha$ & Uniform  & [$-5.0$, $5.0$]  &  - \\
    $r_{eff}$ & Log-uniform  & [$0.005$, $30.0$]  &  $\mu$m

    \end{tabular} 
    \tablefoot{$X_i$ represents the volume mixing ratio abundance of the molecule, $i$.}
    \end{spacing}
    \end{table}

\section{Results and discussion}\label{sec:Results}

In this section, we present the results obtained for the different sets of retrievals. As already discussed, these include three main different approaches: (1) full spectral range; (2) exclusion of the saturated 0.69 - 1.91 $\mu$m range; (3) same as (2) but using a number of cloud extinction models. To display the results, we assume that, following the law of large numbers, the most likely value for each parameter is the median of the marginalised posterior probability distribution. Similarly, the error bars are computed as 1-$\sigma$ deviations from the median.

\subsection{Impact of data reduction on retrieved parameters}\label{subsec:DataCalibrationComparison}

\begin{figure*}
\centering
\includegraphics[width=0.88\hsize]{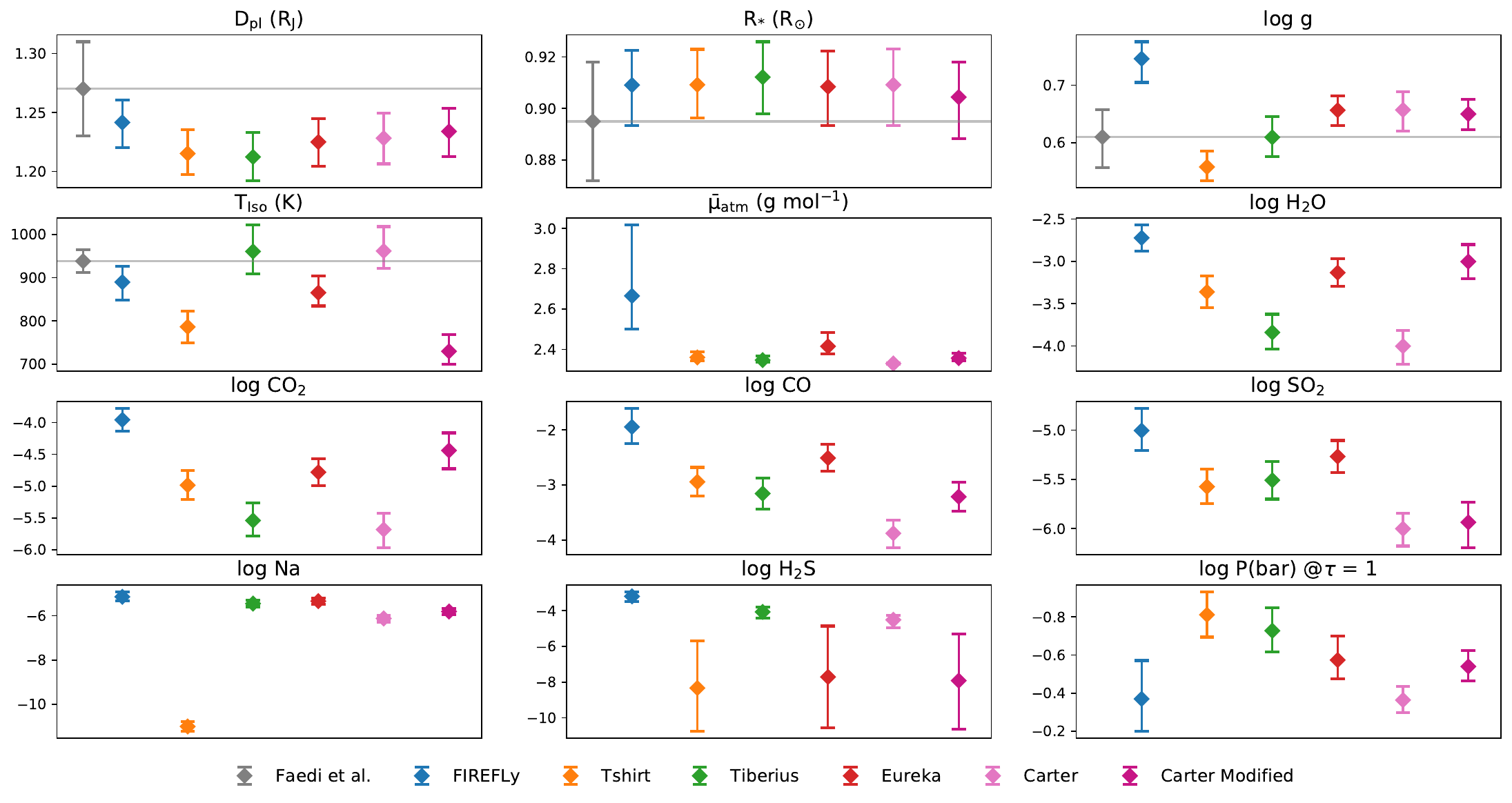}
\caption{Comparison of the retrieval results obtained when using the different spectra as input (shown in different colours). See Sect. \ref{sec:Data} for a full reference of the spectra. The gray values represent the reference values from \cite{Faedi2011}. The planetary radius at 100 mbar, the mean atmospheric molecular weight and the aerosol opacity at 1 $\mu$m were computed from actual outputs and not directly retrieved.}
          \label{FigResultsPipelines}
\end{figure*}

Figure \ref{FigResultsPipelines} shows the resulting parameters (y-axis) for each of the input spectral data (x-axis) mentioned in Sect. \ref{sec:Data}. This set of parameters includes those that physically describe the planet (planetary radius referenced to the 100 mbar pressure level, star radius, planetary gravity, and isothermal temperature), its composition (mean atmospheric molecular weight, abundances for $\rm{H_2O}$, $\rm{CO_2}$ $\rm{CO}$, $\rm{SO_2}$, $\rm{Na}$ and $\rm{H_2S}$), and the aerosols' opacity at 1 $\mu$m, expressed as the pressure level at which the cloud reaches an optical depth $\tau = 1$ in nadir-viewing geometry. For the full analysis, we refer to the corner plots shown in Figs. \ref{FigCornerPlotFIREFLy} to \ref{FigCornerPlotCarterDil} (black lines).

It must be noted that we are including some parameters that are directly used as inputs for the model and some others that are computed afterwards from the outputs, as they have a physical meaning that can provide valuable insights into the best-fitting results. The planetary radius at 100 mbar is calculated combining the retrieved planetary radius at the reference pressure of 10 bar and the atmospheric scale height of the atmosphere. We show the results at this pressure level for an easier comparison with the bibliography. The mean atmospheric weight was calculated using all the fixed and retrieved gases from the modelled atmosphere. Finally, the pressure level at which the cloud reaches $\tau = 1$ in nadir geometry was determined by computing the optical depth at each atmospheric layer using the retrieved cloud abundance. This parameter indicates the altitude at which the atmosphere becomes optically thick with the sole contribution of the aerosols. As we are using a uniform cloud vertical profile, the bigger the value of the pressure level, the thinner the cloud, as it needs more depth to reach the same opacity.

As a reference, we plotted the values of the planetary diameter, the star radius and the gravity of the planet obtained by \cite{Faedi2011} at the discovery of the planet. We also plotted the stratospheric temperature \citep{Barstow2017} computed from the equilibrium temperature of \cite{Faedi2011} because we are mostly sensitive to the upper atmospheric levels \citep{Roy2025}. We note that the stratospheric temperature is always lower than the equilibrium temperature.

As a first conclusion from this inspection, we see that there are clear discrepancies among the retrieved parameters obtained from each data reduction process. Some of them show deviations higher than the retrieved error bars. We go on to analyse the details of some of the most significant differences below.

Both the planetary diameter and the star radius show high compatibility among all the retrieved values and with the reference value, which makes them robust inferences from all modelling efforts. However, the FIREFLy calibration throws a substantially higher value for the gravity, not compatible with any of the other computed values. Regarding the atmospheric temperature, there is also a substantial spread seen across the results, as values differ in some 300K from the hottest (Tiberius) to the coldest (Carter-modified) model. In this case, both Carter-modified and Tshirt appear to be much cooler than expected for the planet \citealp{Faedi2011}, while the rest of calibrations provide models that are compatible with previous values, with all uncertainties taken into account.

Regarding the molecular abundances, there are also big differences. In some cases, the deviations are even bigger than two orders of magnitude. In the particular case of FIREFLy, abundances are systematically higher, thus resulting in an anomalously high mean molecular weight, possibly incompatible with what we might expect for this planet. This could be related to the scale height and gravity anomalies for these data, as already discussed.

When inspecting abundances based on single, narrow absorption bands, such as Na, we also found substantial differences. In this case, it is obvious that retrievals are heavily affected by the spectral grids chosen during data reduction and/or for our simulations. While this is common sense, a word of caution should be raised here for these kinds of absorption features, as the spectral resolution could lead to inconsistent values.

There is also a very interesting case seen for the $\rm{H_2S}$ retrieval. Half of the retrievals (FIREFLy, Tiberius, and Carter) have the capacity to determine its abundance to certain precision, while the other half (Tshirt, Eureka, and Carter-modified) are compatible with no $\rm{H_2S}$ at all in the atmosphere and do not require its presence. While this kind of retrieval is not the ideal way to detect such narrow-banded species, the inclusion or exclusion of certain species is heavily influenced by the calibration process as well.

Finally, there are also differences in the retrieved opacity of the cloud. While the differences are significant, we have found a physically common ground for most of the cases, as they place the limit of an aerosol, optically thick atmosphere for pressure levels at some 100s mbar.

Thus, in short, we find that there is a clear effect of the data reduction process in the model parameters that we retrieved. That can be traced even to the most basic physical parameters of the planet (e.g. gravity or mean molecular weight) and undoubtedly affect the abundances of many species. Even if we can correct the most simple issues by wisely choosing an adequate spectral grid, there is still an uncertainty of at least one order of magnitude that remains. This encourages to always take the values for a given retrieval within the overall context, including the process followed to extract the spectra from the original data.

\subsection{The effect of the saturated region}\label{subsec:ResultsSaturatedRegion}

\begin{figure*}
\centering
\includegraphics[width=0.88\hsize]{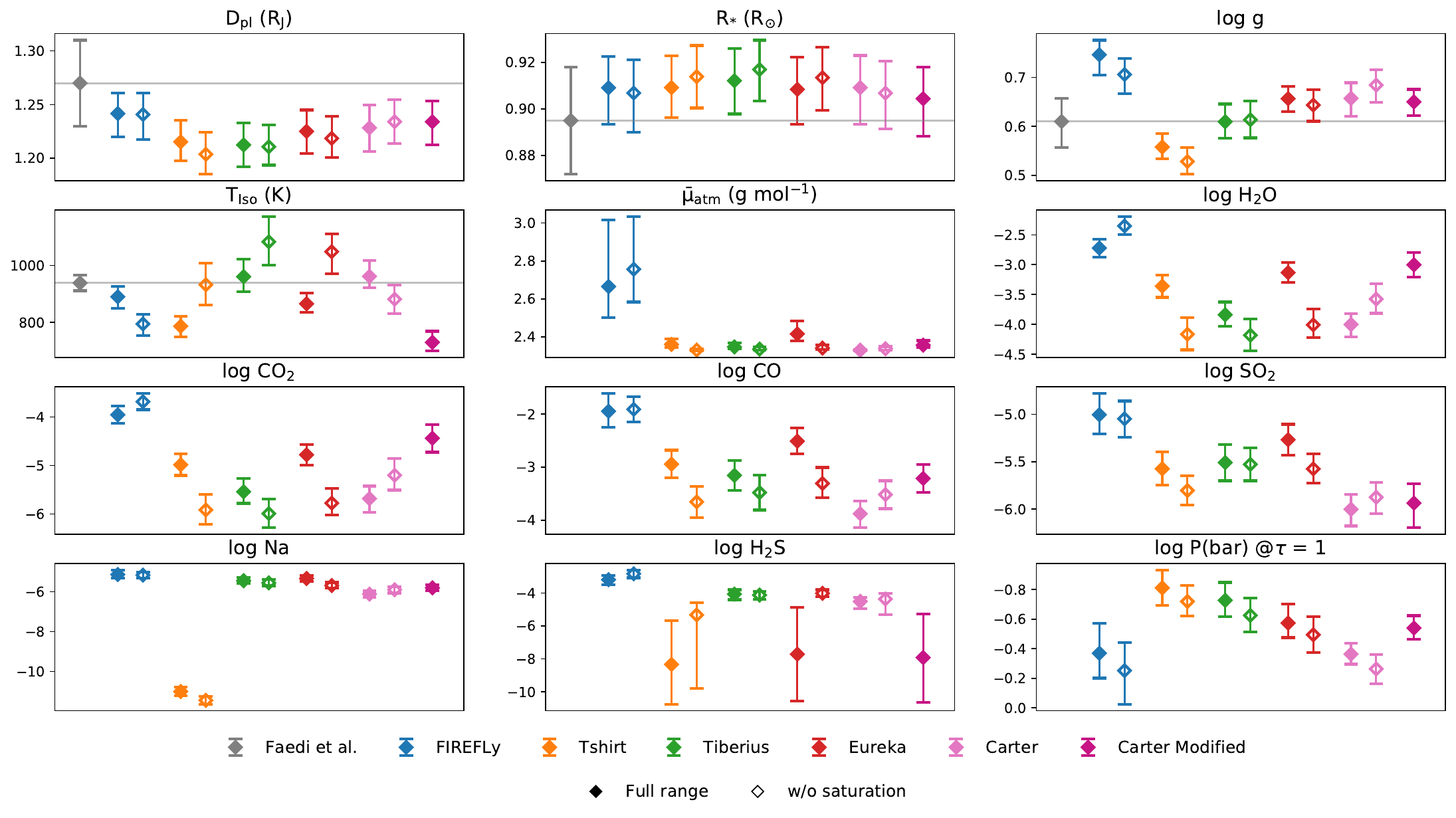}
\caption{Comparison of the results obtained when using the different spectra in retrievals (shown in different colours). See Sect. \ref{sec:Data} for a full reference of the spectra. The filled points correspond to the spectra covering the full spectra range and the unfilled ones to the spectra with the saturated spectral range removed. The gray values represent the reference values from \cite{Faedi2011}. The planetary radius at 100 mbar, the mean atmospheric molecular weight and the aerosols opacity at 1 $\mu$m were computed from actual outputs and not directly retrieved.}
          \label{FigResultsNoSat}
\end{figure*}

We show in Fig. \ref{FigResultsNoSat} the same information as in Fig. \ref{FigResultsPipelines}, but adding the results obtained from the retrievals excluding the saturated region of the spectra. We note that without the saturated region, Carter and Carter-modified are identical and only Carter results were computed at this set of retrievals. For the full analysis, we refer to the corner plots shown in Figs. \ref{FigCornerPlotFIREFLy}-\ref{FigCornerPlotCarterDil} (coloured lines).

The first thing worth highlighting is the fact that the planetary diameter, star radius, and the gravity of the planet do not suffer relevant variations. There are slight differences in the present analysis, but always smaller than the error bars. A similar behaviour was achieved for the aerosol extinction; even if we found some decay in the cloud opacity, it was always compatible with previous results.

\begin{figure}
\centering
\includegraphics[width=0.9\hsize]{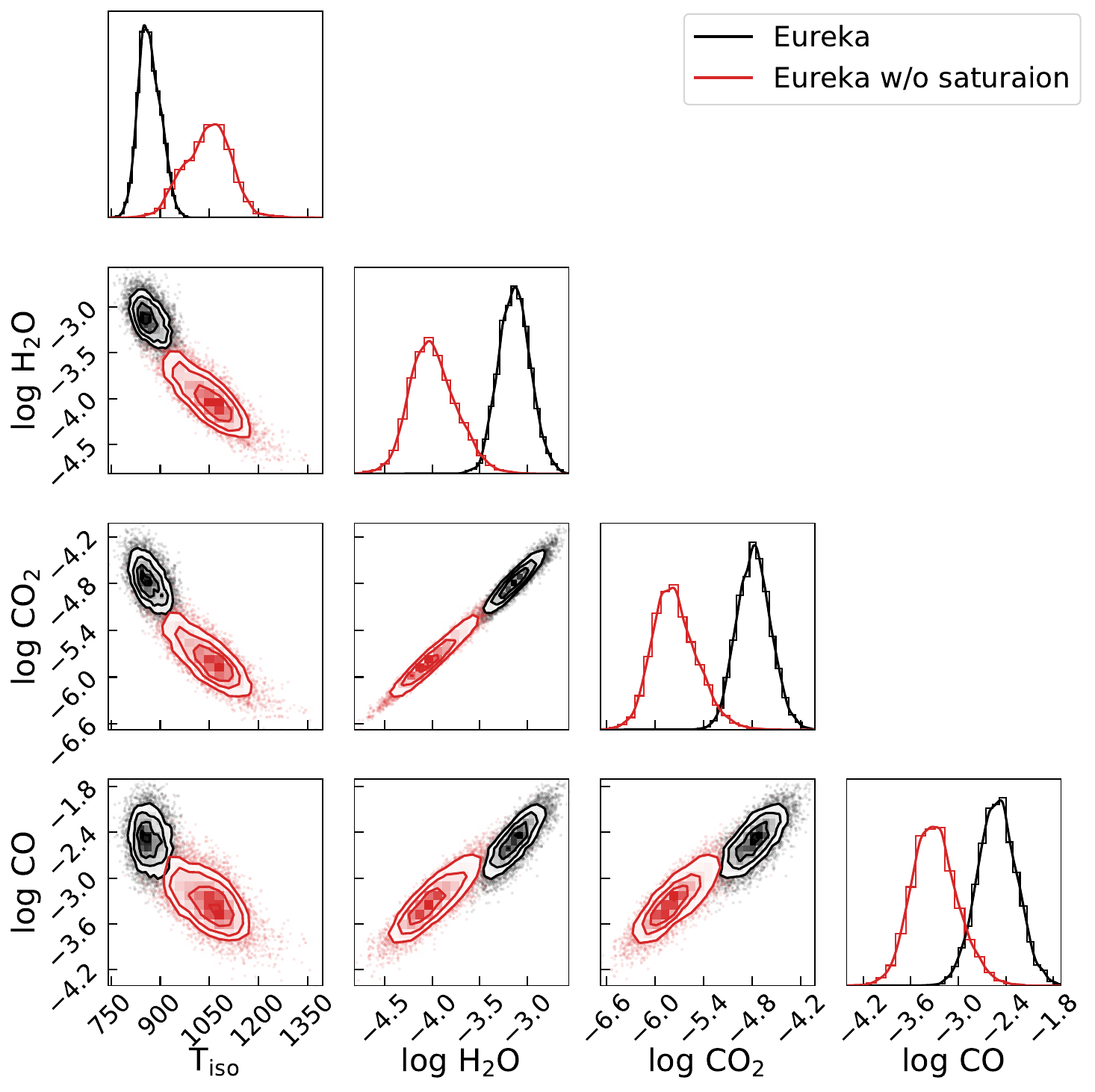}
\caption{Partial corner plot extracted from Fig. \ref{FigCornerPlotEureka}, using Eureka spectrum as the input for the retrievals. Black areas are used for the results including the whole spectral range, while red is used for those excluding the saturated 0.69 - 1.91 $\mu$m range. The temperature shows negative correlations with molecular abundances, while the abundances of $\rm{H_2O}$, $\rm{CO_2}$ and $\rm{CO}$ are positively correlated with each other.}
          \label{FigCornerExtraction}
\end{figure}

However, there are significant differences for the molecular abundances and temperatures, mainly in line with two opposite trends. On the one hand, for the FIREFLy and Carter spectra, the temperature decreases while the molecular abundances are increased. On the other hand, for the rest of the spectra, the temperature increases significantly while the molecular abundances decrease considerably; in some cases, up to an order of magnitude. This anti-correlation can be clearly identified in the mentioned corner plots (see e.g. Figs. \ref{FigCornerExtraction} or \ref{FigCornerPlotEureka}). While it was also present when using the full spectral range, removing the saturated regions increased it substantially. This anti-correlation may be based in the increase of scale height for a hotter atmosphere, which requires us to decrease gaseous abundances for the same extinction to happen.

There are two main aspects that must be considered regarding the saturated region. First of all, it covers mainly some $\rm{H_2O}$ absorption features (see Fig. 4 from \citealp{Rustamkulov2023}). Without the saturated region, the $\rm{H_2O}$ abundance retrieval relies heavily on the absorption peak at 2.8 $\mu$m, which is shared with $\rm{CO_2}$ and $\rm{H_2S}$ absorption, and on the continuum at the red end of the spectra, which is dominated by $\rm{CO}$. This fact increases the degeneracy between these parameters, along with temperature (as summarised in Fig. \ref{FigCornerExtraction}), which leads to lower accuracy when constraining their values. On the other hand, $\rm{SO_2}$ and $\rm{Na}$ are mostly unaffected by the removal of the saturated region. This supports the idea of removing the saturated region, as its inclusion seems to be a substantial source of dispersion for the different spectral data.

It is important to highlight the increased homogeneity of the results from the retrievals without the saturated region, relative to the case presented in the preceding section. Results that diverged in terms of more than an order of magnitude, when using the full range of the spectra, differ by less than half of that when the saturated region is removed.

In summary, taking care of saturation provides a better constrain on parameter degeneracies, but this comes at the cost of increasing result dispersion, depending on the exact procedures for the data calibration. While some prescriptions for the recovery of the saturated region are probably better or more sophisticated than others, comparing calibrations that include this particular region of the spectrum increases the dispersion of the retrieval results, particularly regarding the molecular abundances.

\subsection{Cloud extinction model effect}\label{subsec:ResultsClouds}

\begin{figure*}
\centering
\includegraphics[width=0.88\hsize]{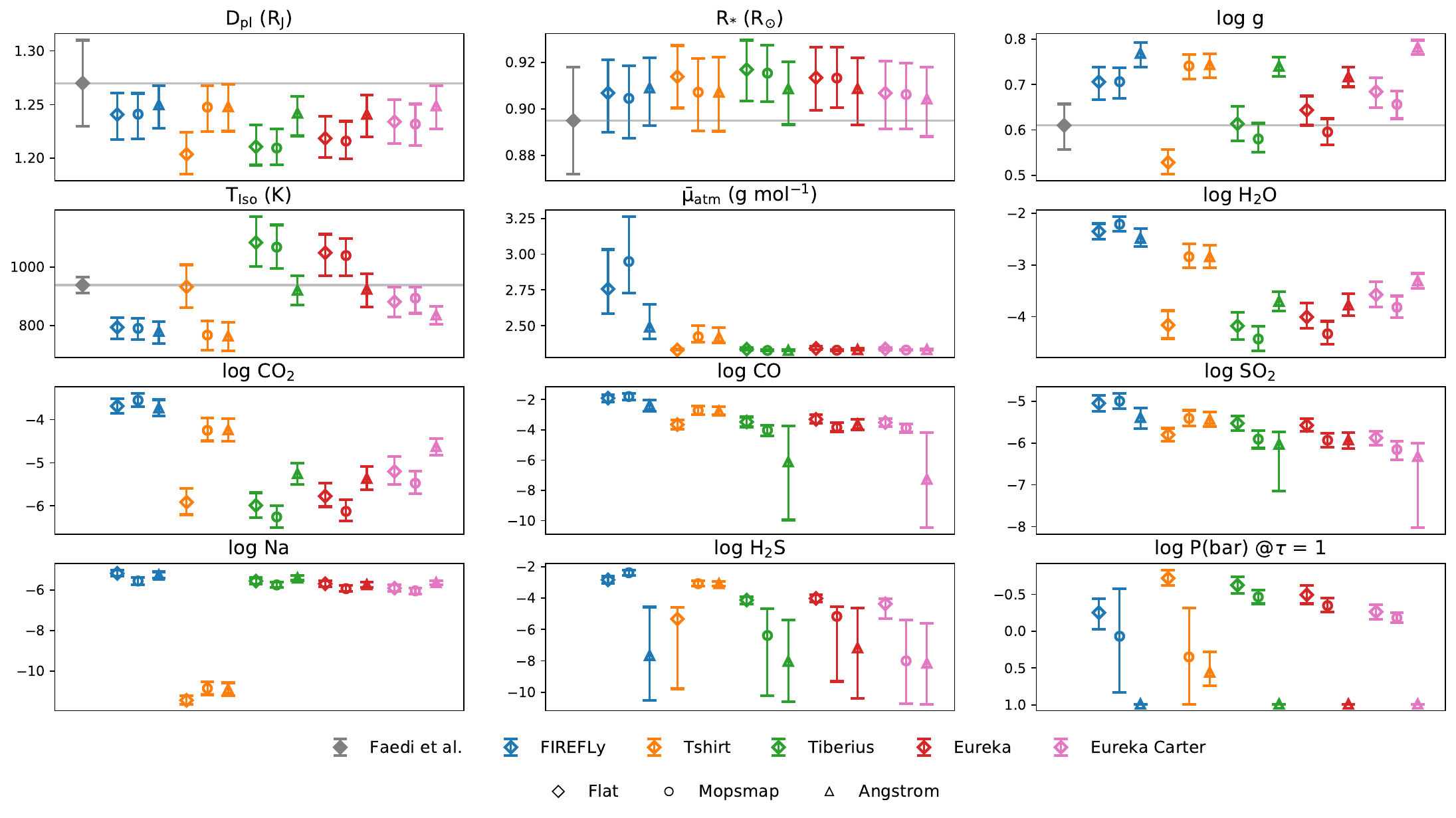}
\caption{Comparison of the results obtained when using the different spectra excluding the saturated region in retrievals. Each colour corresponds to a different input spectral data (see Sect. \ref{sec:Data} for a full reference of the spectra) and each marker shape correspond to a different cloud extinction parametrisation model. The gray values represent the reference values from \cite{Faedi2011}. The planetary radius at 100 mbar, the mean atmospheric molecular weight and the aerosols opacity at 1 $\mu$m were computed from actual outputs and not directly retrieved.}
          \label{FigResultsClouds}
\end{figure*}

Figure \ref{FigResultsClouds} is similar to Fig. \ref{FigResultsNoSat} but showing different cloud extinction parametrisations. Once we demonstrated that including the saturated region in the retrievals introduces a certain level of noise, we then repeated the retrievals excluding the saturation from the spectral data. The `flat' cloud extinction model is the same as in Fig \ref{FigResultsNoSat}. In this figure, we include the results obtained when using the MOPSMAP database and the $\textup{\r{A}}$ngström parametrisation. The disparity in the results for a single reference spectrum when using the different cloud extinction parametrisations is (for most of the atmospheric parameters) at the same level as the dispersion obtained when using the flat cloud extinction model for the different data reductions. This highlights that the biases introduced by data reduction processes in the retrieved atmospheric properties are at the same level as those that can be introduced in a modelling stage.

The results shown in Fig. \ref{FigResultsClouds} allow us to extend the discussion about the role that different cloud extinction models play on the retrieved atmospheric properties, as done in \cite{Roy2025}. The Bayes factors obtained when including the cloud parametrisations with respect to the flat extinction model are shown in Table \ref{TabBayesFactor}. Following Jeffrey's scale criteria \citep{Jeffrey, BayesTrotta}, we find that the Bayesian evidence supports a cloud extinction model more complex than the common `flat' assumption. For the FIREFLy, Tiberius, and Eureka spectra, the Mie scattering parametrisation is strongly favoured, while for the T-shirt and Carter spectra, the $\textup{\r{A}}$ngström parametrisation has higher Bayes factor values; however, the differences are not large enough to claim that one parametrisation is statistically favoured over the other.

\begin{table}[]
\caption{Bayes factor values obtained for the cloud extinction models when compared to the flat model for each of the input spectral data without the saturated region.}
\label{TabBayesFactor}
\centering
\begin{tabular}{lcc}
\hline\hline
Data reduction w/o saturation   & $\ln B_{MOPSMAP}$ & $\ln B_{Angstrom}$\\ \hline
FIREFLy             &  \textbf{15.90} & 8.64   \\
T-shirt            & 8.27  & \textbf{9.19} \\
Tiberius      & \textbf{13.58} & 10.04  \\
Eureka!        &  \textbf{16.28} & 8.34 \\
Carter       & 12.49  & \textbf{13.88}

\end{tabular} 
\end{table}

\begin{figure}
\centering
\includegraphics[width=\hsize]{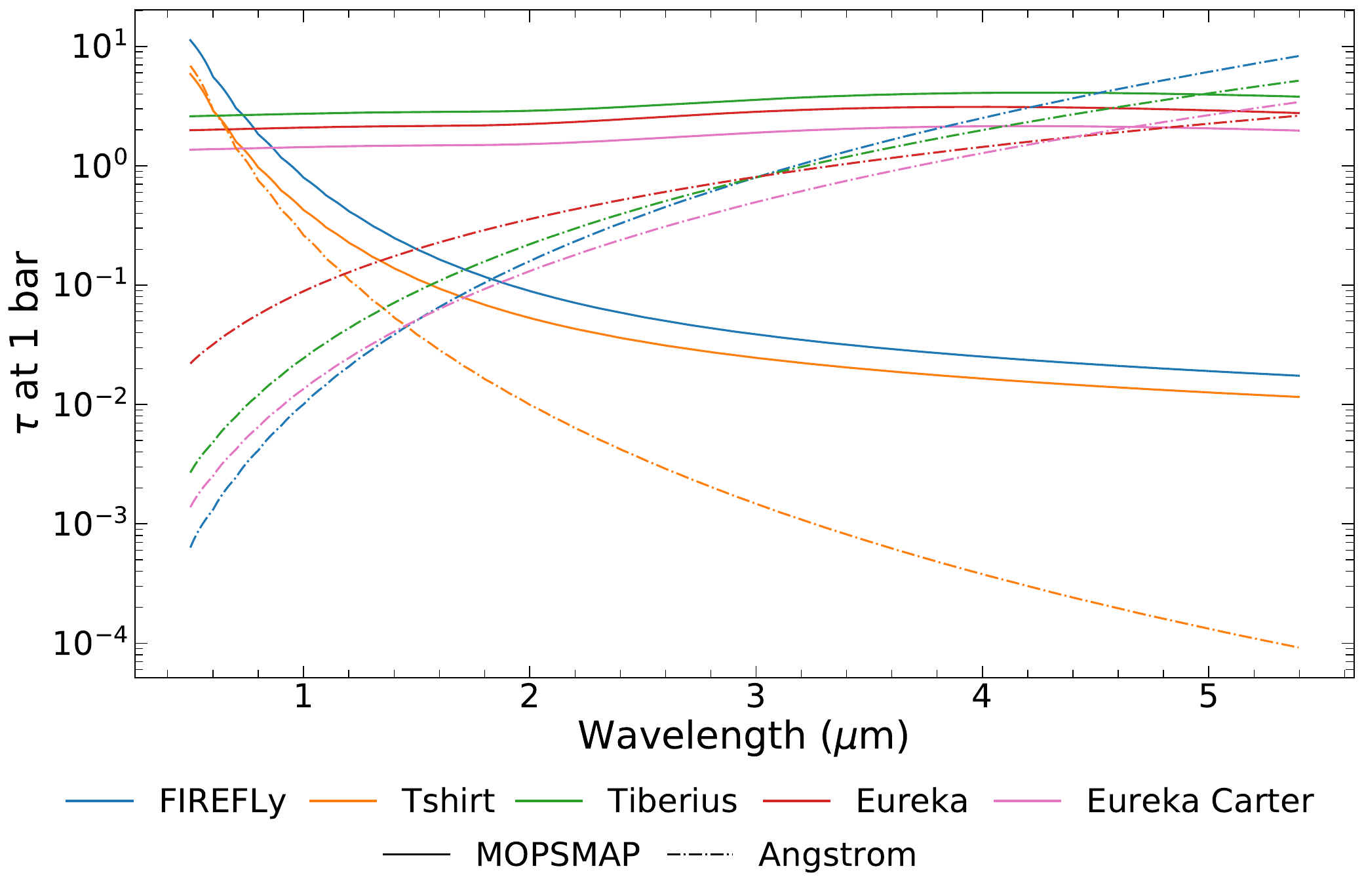}
\caption{Comparison of the retrieved optical depths in nadir geometry along the whole spectral range using cloud extinction parametrisations defined in Sect. \ref{subsec:Clouds}. Each colour corresponds to a different input spectral data. Continuous lines represent optical depth retrieved with the MOPSMAP database and dashed lines represent those computed with the $\textup{\r{A}}$ngström parametrisation.}
          \label{FigOpacitiesClouds}
\end{figure}

The bottom-right panel in Fig. \ref{FigResultsClouds} provides a first glimpse over the properties of the retrieved aerosols for each model, showing their opacity at $\lambda = 1.0$ $\mu$m. Figure \ref{FigOpacitiesClouds}, instead, shows the retrieved cloud optical depth at the 1 bar layer along the whole spectral range for each of the input spectral data and for both aerosol parametrisations. We can identify three main aerosol spectral trends, which is partially correlated with the aerosol opacity, as seen in Fig. \ref{FigResultsClouds}. 

The first group is formed by the clouds defined using MOPSMAP for the Tiberius, Eureka, and Carter spectra. Their overall opacity value is comparable with that retrieved when using flat parametrisations (see Fig. \ref{FigResultsClouds}). However, in terms of the dependence with wavelength, they are not producing flat extinctions anymore. With similar retrieved mean particle size close to $r_{eff} \approx 3.5$ $\mu$m, they reproduce an opacity slightly growing with wavelength. As mentioned in \cite{Roy2025}, the small bump in the spectra, which is statistically favoured with respect to a flat extinction, can hide the contribution from the $\rm{H_2S}$ and prevent its detection.

The second group is formed by the clouds retrieved for the FIREFLy, Tiberius, Eureka, and Carter spectra, but this time using the $\textup{\r{A}}$ngström parametrisation. Figure \ref{FigOpacitiesClouds} shows that for these cases the opacity grows with wavelength. However, with this parametrisation the contrast between the opacity at short and long wavelengths is much higher than that achieved for the first group of clouds. While the optical depth at longer wavelengths is similar to the previous group, the extinction at shorter wavelengths is significantly smaller. We can check Fig. \ref{FigResultsClouds} to see that these models do indeed compensate for the lack of aerosol opacity by increasing the planetary diameter and gravity, while decreasing in temperature. This fact would flatten the spectra and, thus, $\rm{H_2O}$ and $\rm{CO_2}$ abundances would also increase. Despite the low opacity of these clouds, its increase with wavelength also hides the contribution from the $\rm{H_2S}$. In some cases, this effect can even make it difficult to properly constrain the abundances of $\rm{CO}$ and $\rm{SO_2}$, which leave their fingerprints on the longest wavelengths of the spectrum.

Lastly, the third group of cloud models is formed by both clouds (Mie and $\textup{\r{A}}$ngström) retrieved for the Tshirt spectrum and by the MOPSMAP cloud parametrisation for the FIREFLy spectrum, which are very different from previous groups in terms of their spectral properties. Cloud opacities are heavily decreasing with wavelength, corresponding to small particles of sizes $r_{eff} \approx 0.03$ $\mu$m. Inversely to the clouds from the second group, these aerosols produce high optical depths at the shortest wavelengths of the spectrum, while causing almost no extinction at the longest wavelengths. This lower opacity results in the same parameter changes described above: increased diameter, gravity and main molecular abundances, and a decrease in temperature. In this case, as there was no bump towards the longest wavelengths, the retrievals were able to constrain the abundance of $\rm{H_2S}$.

These three aerosol behaviours seem to be related with two different types of retrieved planetary atmospheres by the algorithm. On the one hand, there is the case where the retrieved cloud optical depth is relevant at all the wavelengths of the spectrum. The algorithm evolves towards a planet with lower values of the planetary diameter, but with an extended atmosphere produced by the lower values of the gravity of the planet and the high retrieved temperature. On the other hand, we found a planet with optically thin clouds at some regions of the spectrum. Here, the algorithm retrieves a higher planetary diameter at the reference pressure, but with a more compact atmosphere produced by the high values of the gravity of the planet and its lower atmospheric temperatures. The two solution families are shown in Fig. \ref{FigCornerPlotBestClouds}.

With this information, it is not possible to support which of the two possible atmospheres is statistically favoured. FIREFLy and T-shirt spectra favour the case with a compact atmosphere. On the other hand, Tiberius and Eureka! spectra favour the extended atmosphere obtained when using the MOPSMAP cloud parametrisation. The Carter spectrum shows an intermediate behaviour. Despite achieving cloud opacities similar to those with Tiberius and Eureka!, it simultaneously favours a compact atmosphere. It must be noted, however, that the Bayes factor is just slightly higher than 1, so the evidence is not as clear as in the other cases.

As already mentioned, the parameters defining the extension of the upper atmosphere of the planet are correlated with the abundances of the chemical species present in it. Those retrievals that lead to models with compact atmospheres, in general terms, require higher abundances, especially of $\rm{H_2O}$ and $\rm{CO_2}$. For these models, those with a cloud opacity increasing with wavelength partially hide the contribution from $\rm{CO}$, $\rm{SO_2}$, and $\rm{H_2S}$, even completely masking their contribution in some cases. Nevertheless, those with the strongly decreasing cloud optical depth with wavelength require higher opacities also from those three molecules. This effect translates, specifically for the FIREFLy spectrum, into unexpectedly high atmospheric mean molecular weights. In contrast, the models with more extended atmospheres do not require such high values.

\subsection{Comparison with previous works}\label{subsec:ComparisonBibliography}

Our findings are in general agreement with the conclusions of \cite{Constantinou2023, Powell2024} and \cite{Schleich2025}, specifically finding that there are relevant differences in the retrieved atmospheric properties when using different spectral data as input. We also agree that the most affected parameters are the molecular mixing ratios. Using the Tiberius and Eureka spectra at the 3 - 5 $\mu$m range, \cite{Constantinou2023} found differences in the retrieved abundances of 1 dex or more. Including other spectra, as FIREFLy or Carter, we find that these differences can be even bigger. \cite{Schleich2025} claimed that no other parameters in their retrievals showed any significant variations under perturbations of the transmission spectrum. Nevertheless, \cite{Constantinou2023} found how PT profiles could differ up to 2 $\sigma$ from each other, which is similar to the results presented in the present work. In fact, for some cases we could find even bigger differences. We note that we are using the full NIRSpec/PRISM range, compared to the limited range in \cite{Schleich2025} and the use of HST data in \cite{Constantinou2023}. Then, for the log $g$ parameter, which is usually fixed for the retrievals in previous works, we also obtained substantial differences in its retrieved values. In some cases, especially when including different cloud extinction parametrisations, we obtained substantially higher values not compatible with the previous measurements from \cite{Faedi2011} or \cite{Mancini2018}, as also reported by \cite{Lueber2024} using the NIRSpec/PRISM data.

When removing the saturated region of our input spectra, we get data with reliable spectral points in the NIR and mid-wavelength IR spectral regions. Using these data allows us to reach a higher homogeneity between our results in the retrievals, at the cost of an increased parameter degeneracy. \cite{Constantinou2023} suggested that to obtain accurate abundance estimates with JWST 3 - 5 $\mu$m data, complementary observations at shorter wavelengths are needed. Combining their Tiberius and Eureka spectra with the 0.8-1.7 $\mu$m data from HST, they also managed to obtain a better agreement between the retrieved mixing ratios.

Regarding the role of aerosols in the transmission spectrum, \cite{Constantinou2023} also found evidence suggesting that JWST data have enough potential to constrain the cloud properties. However, their results do not clearly favour any particular condensate or spectral behaviour and they also depend on the data reduction adopted. They obtained aerosols with significant opacity at the wavelengths of HST/WFC3, but with an opacity window at the 3-5 $\mu$m range of NIRSpec/PRISM that could be compatible with those from the third group aerosol behaviours (mentioned in Sect. \ref{subsec:ResultsClouds}).

Lastly, there is a huge range of molecular abundances in the retrievals published so far. For example, \cite{Fisher2024} showed that there is a big range of $\rm{H_2O}$ abundances that are consistent with the data. This is indeed the case for our results: our highest values are compatible with those obtained by \cite{Ahrer2023ERSCO2, Powell2024} or \cite{Fisher2024} with the different JWST instruments during the ERS; our lowest values are in agreement with those reported by \cite{Constantinou2023} using NIRSpec or by \cite{Pinhas2019} using HST and Spitzer. For the other molecular species there are not as many references (see e.g. \citealt{Ahrer2023ERSCO2} or \citealt{Constantinou2023}), but the conclusions seem similar to those found for $\rm{H_2O}$. In summary, the dispersion of values agrees with our results when only considering calibration, while not taking into account other modelling or observing factors.

\section{Conclusion}\label{sec:Conclusions}

The main conclusion of this work is that data reduction process of the JWST observations plays a substantial role for atmospheric retrievals. This is consistent with previous results from other instruments, as \cite{Powell2024} did for MIRI observations. For a set of six different data calibrations of the NIRSpec/PRISM observations of WASP-39b, we find that most of the retrieved atmospheric parameters are heavily affected by the spectrum that has been used as input. Physical parameters such as planetary gravity or temperature show relevant differences among the different retrievals that could lead to different atmospheres. The abundances of the chemical species are heavily influenced by the reference spectrum, showing deviations up to 2 dex in some cases. Thus, it is mandatory to always interpret the values from a given retrieval in the context of the process followed to extract the spectra from the original data.

For the case of WASP-39b, one of the crucial steps during the data reduction process was the recovery of the saturated region of the spectrum. We focus here on the JWST data and the efforts to minimise or remove such saturation. When comparing with retrievals that exclude this spectral region, we find that it introduces more dispersion in the finally retrieved atmospheric parameters, mostly in the abundances of the chemical species of the atmosphere. However, the cost of reducing retrieval dispersion increases the degeneracy between model parameters, as expected from reducing the spectral range. Ideally, a proper correction of the saturated region or, alternatively, a reliable source of information for the same range would be the only solution for removing degeneracies. Meanwhile, combining poor corrections with precise data will just increase the noise for the retrieved range of parameters.

An additional conclusion of this work is that the role of the data processing choices in retrievals is comparable to that of the assumptions and decisions made during the modelling stage. The differences in the retrieved results introduced by choosing different input data is of the same order, if not bigger, than employing different cloud parametrisations when modelling the atmosphere. In fact, both issues are correlated. Including more complex cloud parametrisation is always statistically supported for all the spectral data analysed here, which is a relevant result by itself. However, we show in this work that the FIREFLy and Tshirt spectra, retrievals favour a reduced cloud opacity at some regions of the spectrum together with a compact atmosphere. On the other hand, retrievals with Eureka or Tiberius spectra favour a relevant cloud optical depth along the whole spectral range, which result in atmospheres with higher vertical extensions.

When analysing the retrieved results from each calibration and modelling effort, we find that there are a few possible atmospheres that reproduce the data similarly, but based on quite different versions of the planet. These families or solutions may point to hotter or colder atmospheres, extended or concentrated, and even clear or cloudy at most wavelengths. Leaving aside that some of these combinations may be discarded taking into account further information, present or future, we are faced again with the uncertainty of having to interpret all the data within the context of the reduction process and not just that of the observations.

All in all, a sustained effort to produce reliable calibrations is required. Such an ideal calibration should be in agreement with those provided by other JWST instruments and configurations, as well as with other sources of data, particularly if they complement the spectral range or resolution. While JWST data clearly have the capability to provide information on aspects of the atmosphere, that have not been studied in much detail so far (such as aerosol properties), this still requires an additional effort to achieve an agreement in terms of a general calibration of the data.

\begin{acknowledgements}
This work was supported by Grupos Gobierno Vasco IT1742-22. It has also been supported by grant PID2023‐149055NB‐C31 funded by MICIU/AEI/10.13039/501100011033 and FEDER, UE. J. Roy-Perez acknowledges a PhD scholarship from UPV/EHU. This work is based on observations made with the NASA/ESA/CSA James Webb Space Telescope. The data were obtained from the Mikulski Archive for Space Telescopes at the Space Telescope Science Institute, which is operated by the Association of Universities for Research in Astronomy, Inc., under NASA contract NAS 5-03127 for JWST. These observations are associated with program JWST-ERS-01366. The data used in this paper are associated with JWST program ERS 1366, available from the Mikulski Archive for Space Telescopes (\url{https://mast.stsci.edu}). The authors acknowledge the ERS team for developing their observing program with a zero-exclusive-access period. We thank Dr. G. Villanueva and the PSG team for the development of the tool and the support provided to the users. We would also like to thank E. Ahrer and D. Christie for the discussions during the early stages of this work.
      
\end{acknowledgements}
\bibliographystyle{aa}
\bibliography{aa58193-25}

\clearpage
\onecolumn
\begin{appendix}

\section{Additional figures}

\begin{figure}[!ht]
\centering
\includegraphics[width=\hsize]{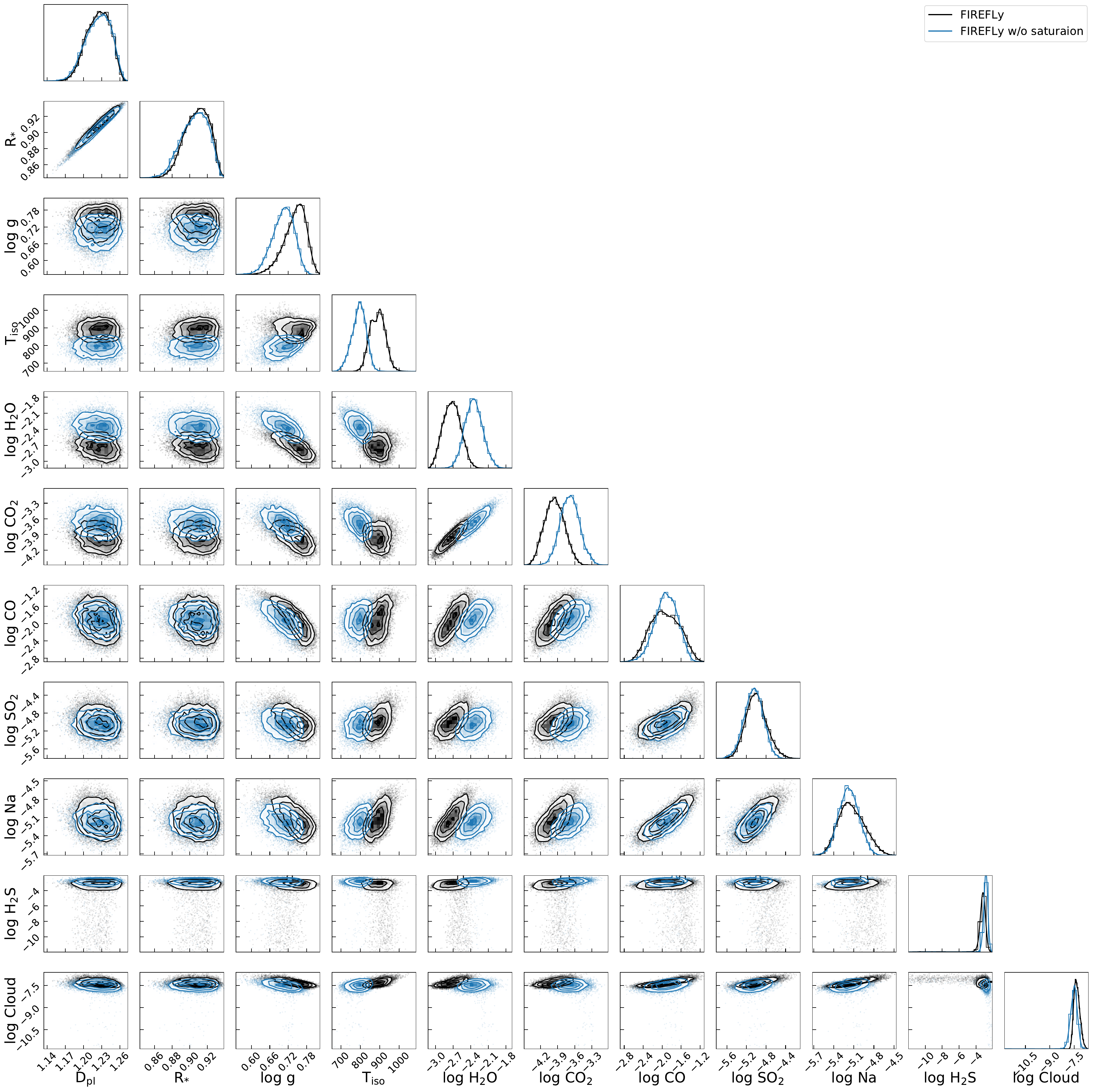}
\caption{Full set of posterior distributions from our nested sampling retrievals performed on the FIREFLy data reduction. Black posteriors refer to retrievals including the whole spectral range of the data and coloured posteriors refer to retrievals excluding the saturated 0.69 - 1.91 $\mu$m range. A simple flat cloud absorption model is considered.}
          \label{FigCornerPlotFIREFLy}
\end{figure}

\begin{figure}
\centering
\includegraphics[width=\hsize]{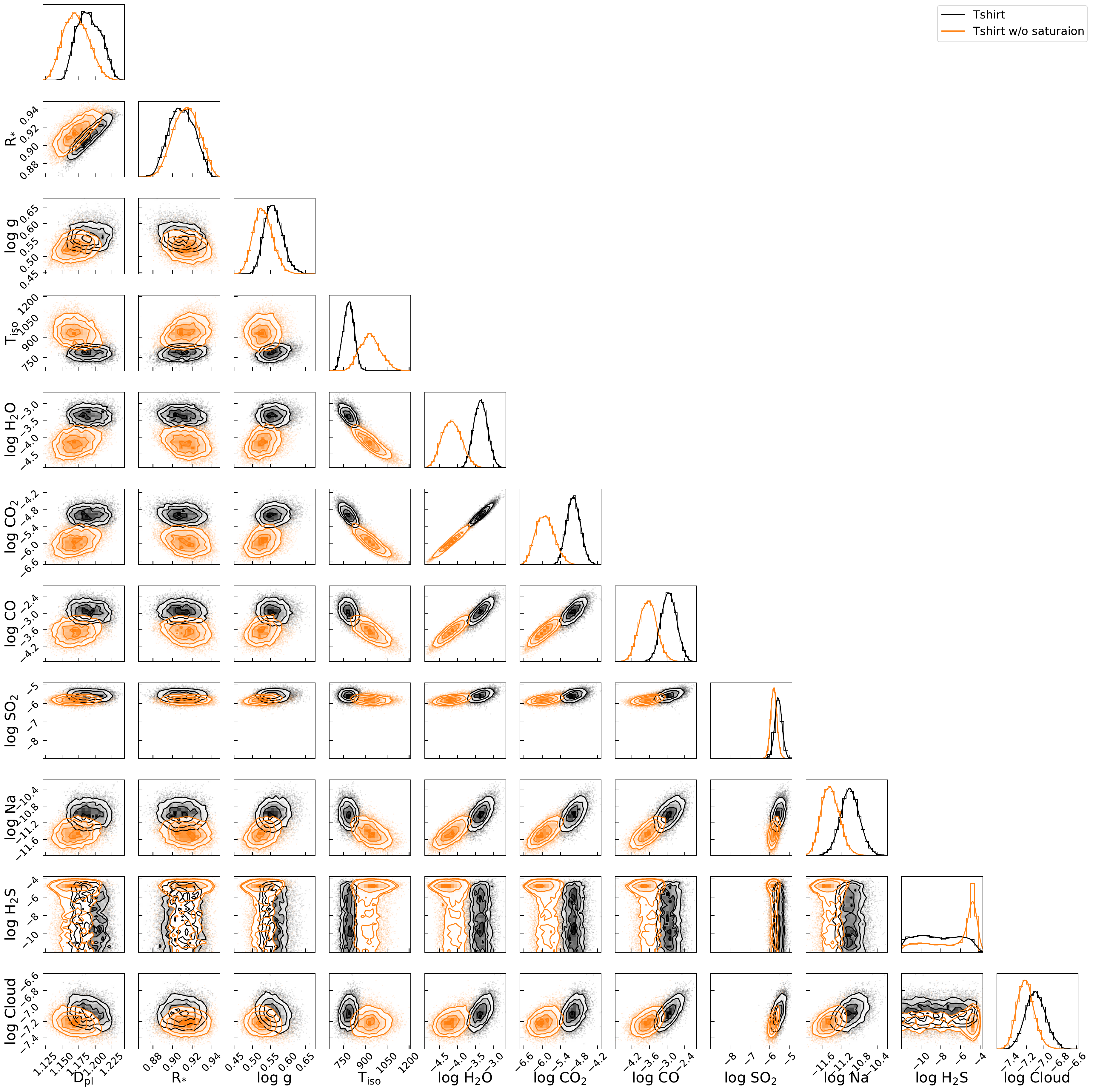}
\caption{Same as Fig. \ref{FigCornerPlotFIREFLy}, but for the Tshirt data reduction.}
          \label{FigCornerPlotTshirt}
\end{figure}

\begin{figure}
\centering
\includegraphics[width=\hsize]{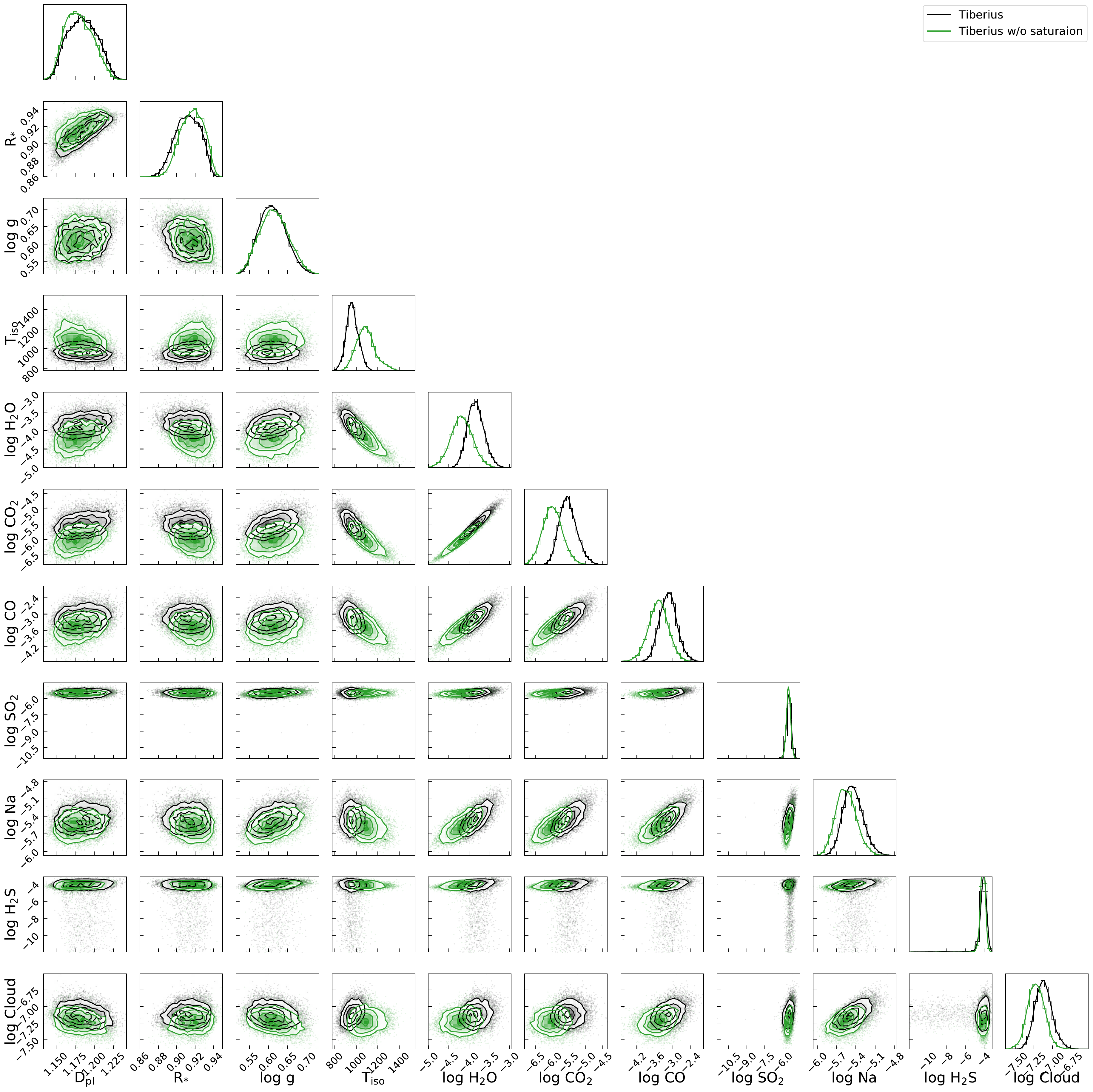}
\caption{Same as Fig. \ref{FigCornerPlotFIREFLy}, but for the Tiberius data reduction.}
          \label{FigCornerPlotTiberius}
\end{figure}

\begin{figure}
\centering
\includegraphics[width=\hsize]{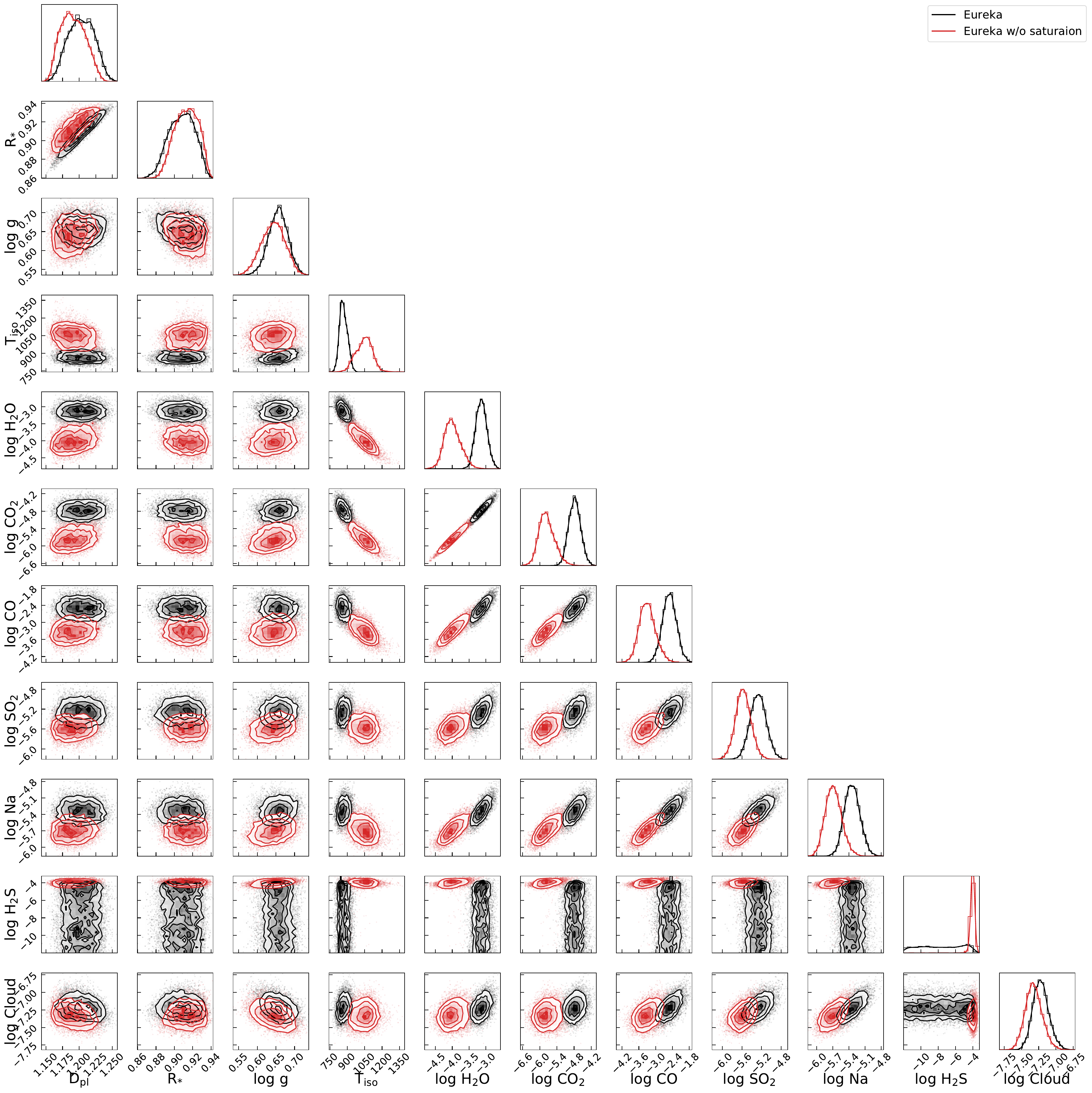}
\caption{Same as Fig. \ref{FigCornerPlotFIREFLy}, but for the Eureka data reduction.}
          \label{FigCornerPlotEureka}
\end{figure}

\begin{figure}
\centering
\includegraphics[width=\hsize]{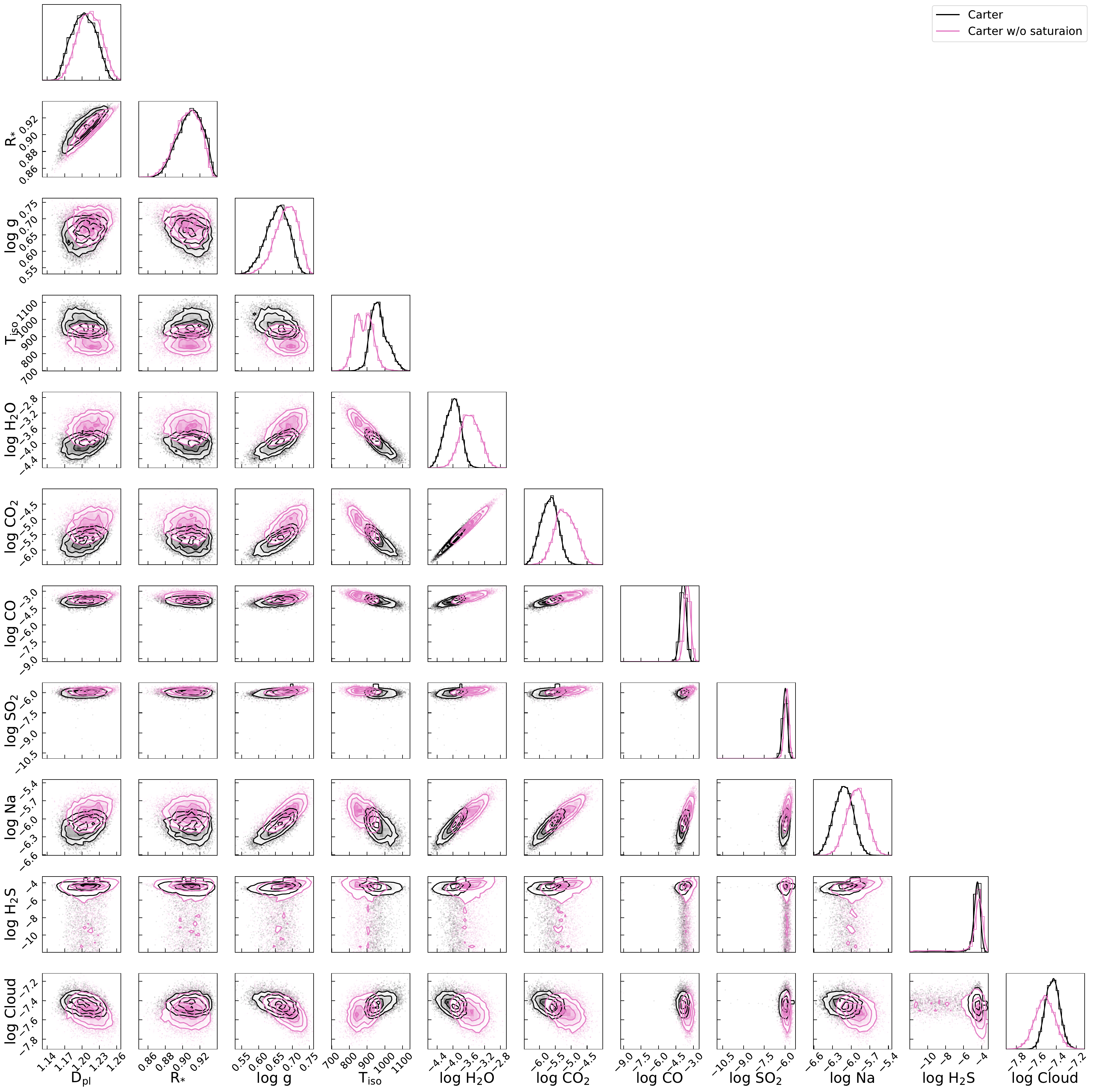}
\caption{Same as Fig. \ref{FigCornerPlotFIREFLy}, but for the Carter data reduction.}
          \label{FigCornerPlotCarter}
\end{figure}

\begin{figure}
\centering
\includegraphics[width=\hsize]{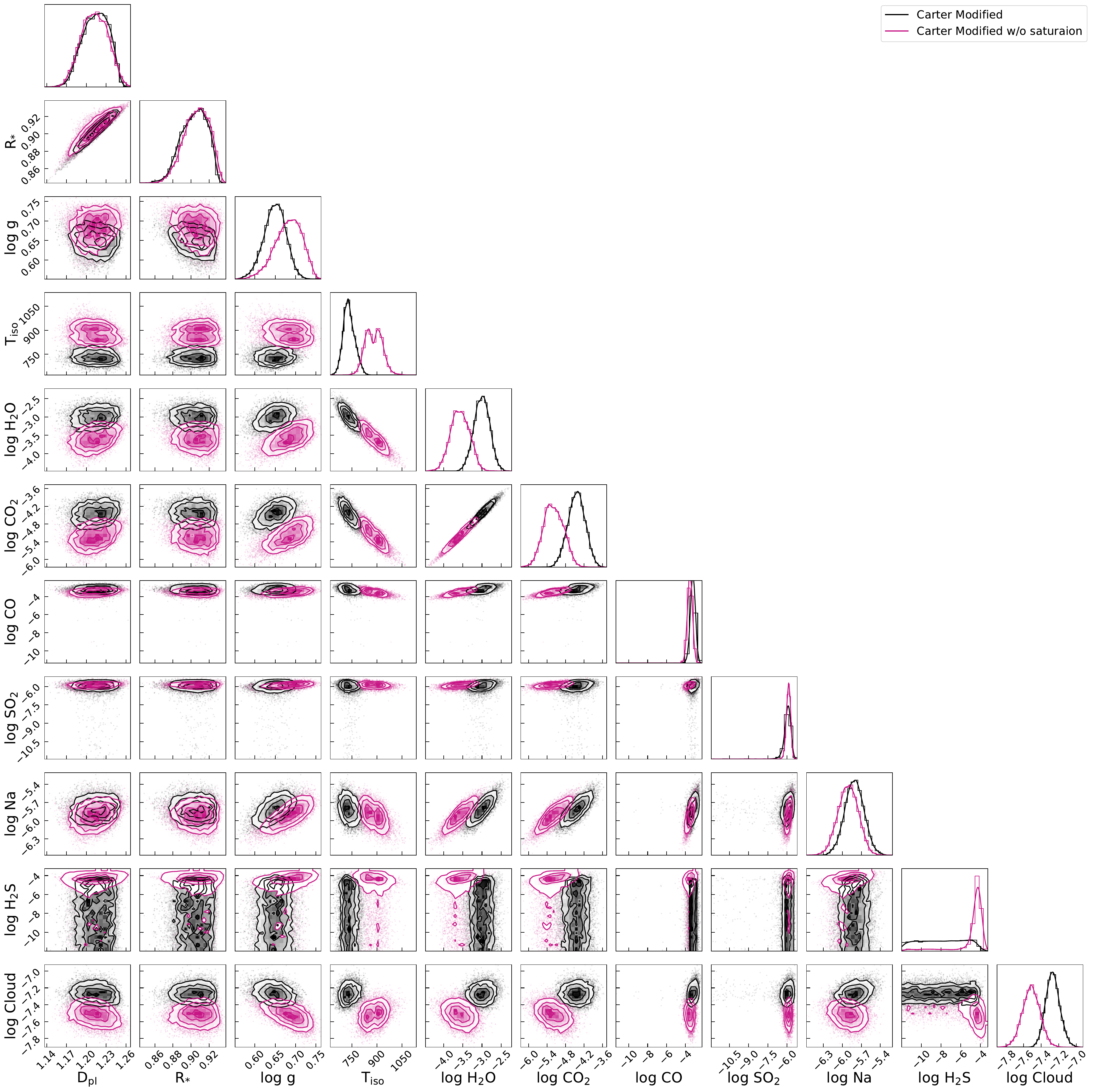}
\caption{Same as Fig. \ref{FigCornerPlotFIREFLy}, but for the Carter-modified data reduction.}
          \label{FigCornerPlotCarterDil}
\end{figure}

\begin{figure}
\centering
\includegraphics[width=\hsize]{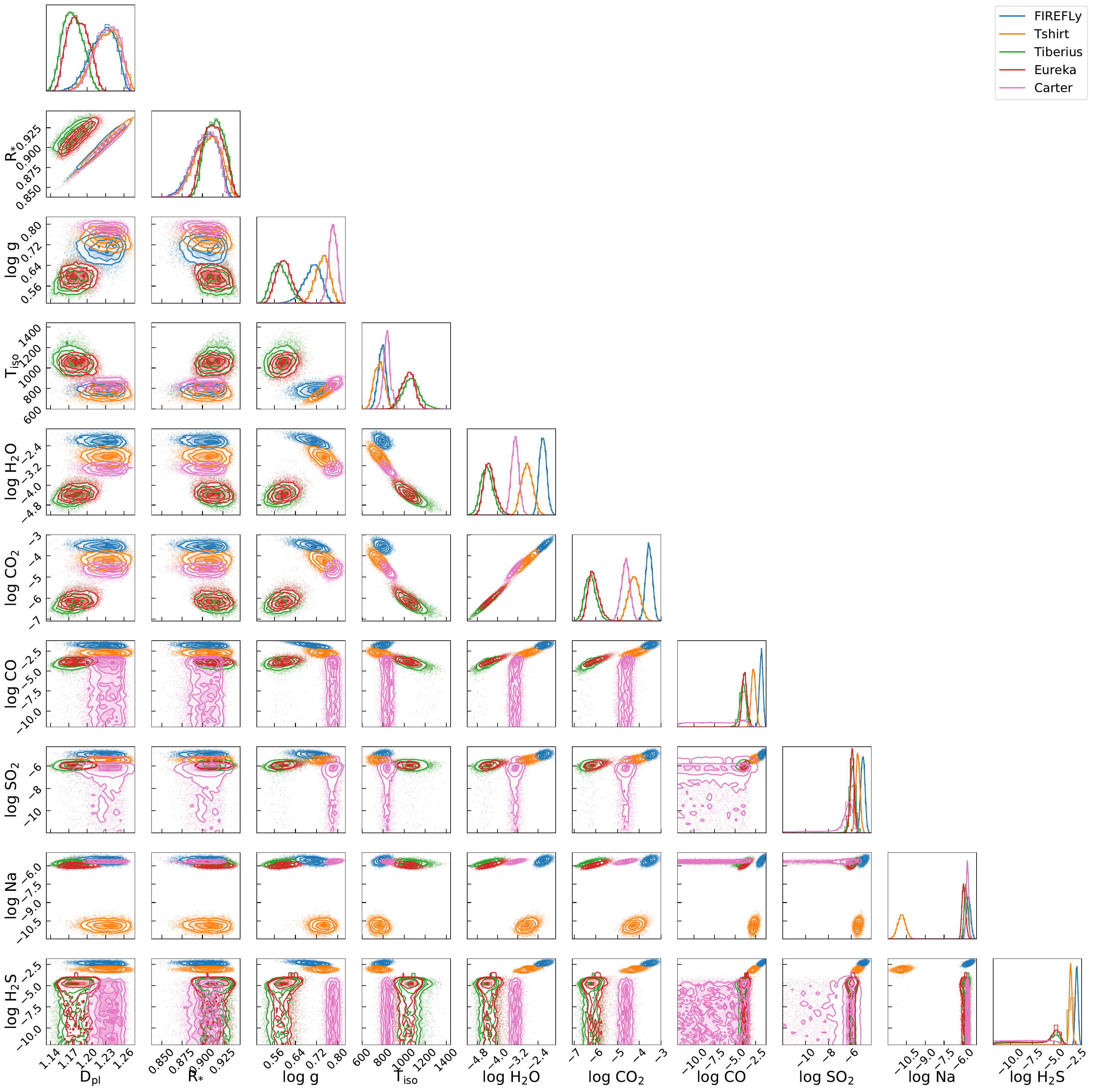}
\caption{Full set of posterior distributions for the best cloud extinction model from our nested sampling retrievals performed on the different data reductions considered in the present work. Retrievals have been performed excluding the saturated 0.69 - 1.91 $\mu$m range from the data.}
          \label{FigCornerPlotBestClouds}
\end{figure}

\end{appendix}

\end{document}